\documentclass[12pt]{article}
\pdfoutput=1

\usepackage{amsmath}
\usepackage{amsfonts}
\usepackage{amscd}
\usepackage{amsthm}
\usepackage{setspace}

\usepackage{graphicx}
\usepackage{authblk}
\usepackage{caption}
\usepackage{ytableau}
\usepackage{mathtools}

\def\Z{\mathbb{Z}}

\def\P{\mathbb{P}}

\def\til{\tilde}

\begin{document}

\begin{titlepage}

\begin{flushright}
KEK-TH-2158
\end{flushright}

\vskip 1cm

\begin{center}

{\bf \Large $\frac{1}{2}$ Calabi--Yau 3-folds, Calabi--Yau 3-folds as double covers, and F-theory with U(1)s}

\vskip 1.2cm

Yusuke Kimura$^1$ 
\vskip 0.6cm
{\it $^1$KEK Theory Center, Institute of Particle and Nuclear Studies, KEK, \\ 1-1 Oho, Tsukuba, Ibaraki 305-0801, Japan}
\vskip 0.4cm
E-mail: kimurayu@post.kek.jp

\vskip 2cm
\abstract{In this study, we introduce a new class of rational elliptic 3-folds, which we refer to as ``1/2 Calabi--Yau 3-folds''. We construct elliptically fibered Calabi--Yau 3-folds by utilizing these rational elliptic 3-folds. The construction yields a novel approach to build elliptically fibered Calabi--Yau 3-folds of various Mordell--Weil ranks. Our construction of Calabi--Yau 3-folds can be considered as a three-dimensional generalization of the operation of gluing pairs of 1/2 K3 surfaces to yield elliptic K3 surfaces. From one to seven $U(1)$s form in six-dimensional $N=1$ F-theory on the constructed Calabi--Yau 3-folds. Seven tensor multiplets arise in these models.}  

\end{center}
\end{titlepage}

\tableofcontents
\section{Introduction}
\par F-theory \cite{Vaf, MV1, MV2} is compactified on spaces that admit a torus fibration, and axiodilaton in type IIB superstrings is identified with the modular parameter of elliptic curves as fibers of the torus fibration. The formulation enables the axiodilaton to exhibit $SL_2(\Z)$ monodromy. 
\par When the compactification space includes two or more independent global sections, $U(1)$ gauge group forms in F-theory \cite{MV2}. Global section is a structure of the torus fibration that is obtained when a point in an elliptic curve as a fiber is selected over each point in the base space and when the chosen point can be moved smoothly throughout over the base of the torus fibration to yield a copy of the base space. When the operation is feasible, genus-one fibration includes a global section, and a genus-one fibration admitting a global section is typically referred to as an elliptic fibration in the F-theory literature \footnote{Recent studies of F-theory models on elliptic fibrations that admit a global section include \cite{MorrisonPark, MPW, BGK, BMPWsection, CKP, BGK1306, CGKP, CKP1307, CKPS, Mizoguchi1403, AL, EKY1410, LSW, CKPT, CGKPS, MP2, MPT1610, BMW2017, CL2017, BMW1706, KimuraMizoguchi, Kimura1802, LRW2018, MizTani2018, Kimura1810, CMPV1811, TT2019, Kimura1902, Kimura1903, EJ1905, LW1905}.}. The rank of the group formed by the set of the global sections of the compactification space -or the ``Mordell--Weil group''- yields the number of $U(1)$ factors forming in F-theory on the space \cite{MV2}. The rank of the ``Mordell--Weil group'' is one less than the number of independent sections, and thus the number of $U(1)$ factors that form in F-theory is one less than the number of the independent global sections. F-theory models in which one or more $U(1)$ factors form are analyzed include \cite{MorrisonPark, BMPWsection, CKP, CGKP, BMPW2013, CKPS, MTsection, MT2014, KMOPR, BGKintfiber, GKK, MPT1610, CKPT, Kimura1802, TT2018, CMPV1811, TT2019, Kimura1908}.

\vspace{5mm}

\par The presence of $U(1)$ aids in explaining some characteristic properties of GUT such as mass hierarchies of quarks and leptons and a suppression of proton decay, and it relates to realization of GUT in the formulation of F-theory. 

\vspace{5mm}

\par The aim of this study is to introduce and study a novel approach to systematically construct six-dimensional (6D) F-theory models wherein various numbers of $U(1)$ factors form. Specifically, we introduce a new method to construct elliptically fibered Calabi--Yau 3-folds with various Mordell--Weil ranks. F-theory on the constructed spaces yields 6D $N=1$ theories with $U(1)$ factors from one to seven. 
\par In this study, we use certain rational 3-folds which we refer to as ``1/2 Calabi--Yau 3-folds'' to construct elliptically fibered Calabi--Yau 3-folds with positive Mordell--Weil ranks. Elliptically fibered Calabi--Yau 3-folds are obtained by considering double covers of the ``1/2 Calabi--Yau 3-folds''. We construct ``1/2 Calabi--Yau 3-folds'' by blowing up $\P^3$ in eight points, and these rational 3-folds naturally have elliptic fibrations as explained in section \ref{section2.1}. The ``1/2 Calabi--Yau 3-folds'' can be viewed as a generalization of ``1/2 K3 surfaces'' \footnote{These elliptic surfaces are also typically referred to as rational elliptic surfaces. (Rational elliptic surfaces correspond to elliptically fibered surfaces obtained by blowing up the complex projective plane $\P^2$ in the nine base points of a cubic pencil.)} to 3-folds. As discussed in section \ref{section2.1} and in appendix \ref{appendixA}, several characteristic properties of 1/2 K3 surfaces include analogs for ``1/2 Calabi--Yau 3-folds'' while other properties of 1/2 K3 surfaces do not directly generalize to these rational elliptic 3-folds. The rank of the Mordell--Weil group and rank of the $ADE$ singularity of every ``1/2 Calabi--Yau 3-fold'' add to 7. This is considerably analogous to a characteristic property of the 1/2 K3 surfaces wherein the ranks of the Mordell--Weil group and the $ADE$ singularity type add to 8. Details of the property of ``1/2 Calabi--Yau 3-fold'' are discussed in section \ref{section2.1} and appendix \ref{appendixA}. We use the property of ``1/2 Calabi--Yau 3-folds'' to construct elliptically fibered Calabi--Yau 3-folds of various Mordell--Weil ranks. We consider a double cover of ``1/2 Calabi--Yau 3-fold'' to yield an elliptically fibered Calabi--Yau 3-fold, and this can be seen as an analog of the operation wherein a pair of 1/2 K3 surfaces are glued to yield an elliptic K3 surface. The base of the elliptically fibered Calabi--Yau 3-fold obtained by taking double cover of ``1/2 Calabi--Yau 3-fold'' is isomorphic to the del Pezzo surface of degree two, and this is isomorphic to the blow-up of seven points on the projective plane $\P^2$. Thus, seven tensor multiplets arise in F-theory on the resulting Calabi--Yau 3-folds. This is discussed in section \ref{section2.1}. 
\par The construction of the aforementioned elliptically fibered Calabi--Yau 3-folds can be useful in analyzing $U(1)$ gauge symmetries forming in 6D $N=1$ F-theory. The construction also yields a novel approach to build families of elliptically fibered Calabi--Yau 3-folds. 

\vspace{5mm}

\par The construction of elliptically fibered Calabi--Yau 3-folds as double covers of ``1/2 Calabi--Yau 3-folds'' to yield 6D F-theory models in this study  can be considered as a generalization of the approach in \cite{Kimura1802, Kimura1903} yielding elliptic K3 surfaces by gluing pairs of 1/2 K3 surfaces to obtain (four-dimensional and eight-dimensional) F-theory models with various $U(1)$ factors, in the sense that taking double covers of ``1/2 Calabi--Yau 3-fold'' corresponds to a higher dimensional analog of the operation of gluing two 1/2 K3 surfaces to yield an elliptic K3 surface. 

\vspace{5mm}

\par \cite{Nak, DG, G} analyzed the structures of the elliptic fibrations of 3-folds.

\vspace{5mm}

\par Recent studies of F-theory model buildings mainly focused on the use of the local models \cite{DWmodel, BHV1, BHV2, DW}. However, it is necessary to analyze global aspects of the geometry of the compactification space to deal with the problems of the early universe including inflation and issues of gravity. The structures of the elliptically fibered Calabi--Yau 3-folds are examined from a global perspective in this study. 

\vspace{5mm}

\par We provide a summary of the obtained results and the construction of 6D $N=1$ F-theory models with multiple U(1) factors in section \ref{sectionsummary}. In section \ref{section2.1}, we discuss the constructions of the ``1/2 Calabi--Yau 3-folds'' and elliptically fibered Calabi--Yau 3-folds as double covers of them. We discuss 6D $N=1$ F-theory compactifications on the resulting Calabi--Yau 3-folds. The resulting Calabi--Yau 3-folds have Mordell--Weil groups of various ranks. $U(1)^n$, $n=1, \ldots, 7$, forms in F-theory on the resulting Calabi--Yau 3-folds depending on the rank of the Mordell--Weil group of the Calabi--Yau 3-fold. We discuss some specific examples of the constructions of Calabi--Yau 3-folds in sections \ref{section2.2}, \ref{section2.3}, \ref{section2.4}, \ref{section2.5}, and \ref{section2.6} as a demonstration of our construction. 

\vspace{5mm}

\par We consider the relation between exceptional gauge group factors and numbers of tensor fields forming in 6D $N=1$ F-theory in section \ref{section3} by comparing 6D F-theory models on the elliptically fibered Calabi--Yau 3-folds that are constructed in this study with the 6D F-theory models constructed in \cite{Kimura1810, Kimura1902}. We also note a possible relation to the swampland conditions. Recent reviews of the swampland criteria are given in \cite{BCV1711, Palti1903}. The notion of the swampland was discussed in \cite{Vafa05, AMNV06, OV06}. The authors in \cite{KSV1905} recently discussed swampland conditions for 6D $N=1$ supergravity theories by considering consistency condition on 2D strings coupled to 6D supergravity theories. Conditions on the possible combinations of distinct gauge groups and matter fields for 6D $N=1$ quantum gravitational theories were discussed in \cite{KT0910, KMT1008, PT1110}. A review of this topic can be found in \cite{Taylor1104}.

\vspace{5mm}

\par A few open problems that remain are discussed in section \ref{section4}.

\vspace{5mm}

\par In appendix \ref{appendixA}, certain details of the constructions of the ``1/2 Calabi--Yau 3-folds'' and elliptically fibered Calabi--Yau 3-folds as their double covers are discussed.

\section{Summary of the results}
\label{sectionsummary}
We provide the summary of the main results obtained in this study. 
\par When one considers three quadrics, $Q_1, Q_2, Q_3$ in $\P^3$, they intersect in eight points. We consider the blow-up of $\P^3$ in the eight intersection points, and this yields what we refer to in this study as ``1/2 Calabi--Yau 3-fold''. The ``1/2 Calabi--Yau 3-folds'' have an elliptic fibration as we will discuss in section \ref{section2.1.1}. The ``1/2 Calabi--Yau 3-folds'' are useful in constructing 6D $N=1$ F-theory models with multiple $U(1)$ factors because these elliptic 3-folds satisfy the following equality:
\begin{equation}
\label{sum 7 in 2.1}
{\rm rk}\, ADE(X) + {\rm rk}\, MW(X) =7.
\end{equation}
$ADE(X)$ denotes the $ADE$ singularity type of 1/2 Calabi--Yau 3-fold $X$, and $MW(X)$ denotes the Mordell--Weil group of 1/2 Calabi--Yau 3-fold $X$. Given this equality, the Mordell--Weil ranks of 1/2 Calabi--Yau 3-fold take values from zero to seven. We focus on the members of the 1/2 Calabi--Yau 3-fold with the Mordell--Weil ranks from one to seven in this study. 
\par We consider double covers of the 1/2 Calabi--Yau 3-folds to yield elliptically fibered Calabi--Yau 3-folds as described in section \ref{section2.1.2}. The equation of double cover is givens as (\ref{CY double cover in 2.1}) in section \ref{section2.1.2}. The base surface of the obtained Calabi--Yau 3-folds is isomorphic to the del Pezzo surface of degree two $dP_7$, and this follows from a method in mathematics. Given that the del Pezzo surface of degree two is obtained by blowing up the projective plane $\P^2$ in seven points, seven tensor multiplets arise in 6D $N=1$ F-theory for the resulting Calabi--Yau 3-folds. 
\par As discussed in section \ref{section2.1.2}, the operation of taking double cover can be viewed as a base change, and thus a global section of the original 1/2 Calabi--Yau 3-fold $X$ lifts to a section of the Calabi--Yau 3-fold $Y$ obtained by considering the double cover via the base change. This implies that the following inequality holds between the Mordell--Weil ranks of the original 1/2 Calabi--Yau 3-fold $X$ and the Calabi--Yau 3-fold $Y$ obtained by taking the double cover:
\begin{equation}
{\rm rk}\, MW(Y) \ge {\rm rk}\, MW(X).
\end{equation}
Furthermore, we expect that the equality actually holds between the two Mordell--Weil ranks as mentioned in section \ref{section2.1.2}. 
\par Thus, elliptically fibered Calabi--Yau 3-folds of Mordell--Weil ranks at least from one to seven are obtained as a result of taking double covers of 1/2 Calabi--Yau 3-folds of Mordell--Weil ranks from one to seven. F-theory compactifications on the resulting Calabi--Yau 3-folds yield 6D $N=1$ theories in which from one to seven $U(1)$ factors form. 
\par When the generic three quadrics $Q_1,Q_2,Q_3$ are selected, 1/2 Calabi--Yau 3-folds obtained by blowing up $\P^3$ in the eight base points have Mordell--Weil rank seven. When we select specific three quadrics $Q_1,Q_2,Q_3$, 1/2 Calabi--Yau 3-folds with Mordell--Weil ranks of six, five, four, and one are also obtained. Seven, six, five, four, and one $U(1)$ factors arise in 6D $N=1$ F-theory models on the Calabi--Yau 3-folds as obtained by considering double covers of the constructed 1/2 Calabi--Yau 3-folds. The explicit examples of 6D F-theory models with multiple $U(1)$ factors are discussed in sections \ref{section2.2}, \ref{section2.3}, \ref{section2.4}, \ref{section2.5}, and \ref{section2.6}.
\par By comparing other 6D $N=1$ F-theory models constructed in \cite{Kimura1810, Kimura1902} with 6D $N=1$ F-theory models constructed in this note, we discuss the possible numbers of the exceptional gauge group factors that can form in 6D $N=1$ F-theory with seven tensor multiplets in section \ref{section3}.

\section{Calabi--Yau 3-folds as double covers of 1/2 Calabi--Yau 3-folds and F-theory with U(1) factors from one to seven}
\label{section2}

\subsection{Construction of Calabi--Yau 3-folds and summary of F-theory compactifications on them}
\label{section2.1}

\subsubsection{1/2 Calabi--Yau 3-folds}
\label{section2.1.1}
\par We construct families of elliptically fibered Calabi--Yau 3-folds by making use of ``1/2 Calabi--Yau 3-folds''. The construction of ``1/2 Calabi--Yau 3-folds'' can be viewed as a higher dimensional generalization of blowing up $\P^2$ in nine points to yield a 1/2 K3 surface. Given that considering a double cover of these rational 3-folds (branched along an appropriate divisor) yields elliptically fibered Calabi--Yau 3-folds, we refer to such rational elliptic 3-folds as ``1/2 Calabi--Yau 3-folds'' in this study.  
\par 1/2 Calabi--Yau 3-folds $X$ satisfy the equality (\ref{sum 7 in 2.1}) as we stated in section \ref{sectionsummary}. The singularity rank and the Mordell--Weil rank of 1/2 Calabi--Yau 3-folds always add to seven and this sum is independent of the complex structure moduli of 1/2 Calabi--Yau 3-folds. Details of the equality including a sketch of a proof is discussed in appendix \ref{appendixA}. The equality (\ref{sum 7 in 2.1}) is quite analogous to the property of 1/2 K3 surface that the ranks of the Mordell--Weil group and the singularity type of 1/2 K3 surface always add to eight. The property (\ref{sum 7 in 2.1}) of 1/2 Calabi--Yau 3-folds enables us to construct elliptically fibered Calabi--Yau 3-folds of various Mordell--Weil ranks.  

\vspace{5mm}

\par We discuss a few more details of the construction and properties of 1/2 Calabi--Yau 3-folds. We select three (homogeneous) quadrics, $Q_1, Q_2, Q_3$, in $\P^3$. The three quadrics intersect in points wherein the number corresponds to the product of their degrees, $2\times 2\times 2$, i.e., they intersect in eight points. Thus, the three quadrics have the simultaneous zeros at the eight points in $\P^3$. We consider blow-up of $\P^3$ in the eight points as the intersection of the three quadrics, and this operation yields a rational 3-fold. Taking the ratio
\begin{equation}
[Q_1:Q_2:Q_3]
\end{equation}
yields a projection onto $\P^2$. The fiber of the projection over a point $[a:b:c]$ in $\P^2$ is the complete intersection of two quadrics in $\P^3$ as follows:
\begin{eqnarray}
b\, Q_1-a\, Q_2 & =0 \\ \nonumber
c\, Q_2-b\, Q_3 & =0,
\end{eqnarray}
which is an elliptic curve. Therefore, the projection yields an elliptic fibration, and the resulting rational 3-fold is a rational elliptic 3-fold. We refer to the resulting 3-fold as a 1/2 Calabi--Yau 3-fold. 
\par $\P^2$s that arise from the eight points when the points are blown up yield global sections to the elliptic fibration. This is very analogous to the situation of 1/2 K3 surfaces wherein the $\P^1$s arise from the nine points in $\P^2$ when they are blown up to yield global sections to the elliptic fibration of 1/2 K3 surface. When one of the eight global sections arising from the blown-up eight points is selected as a zero-section, the remaining seven sections generate the Mordell--Weil group. Therefore, when the three quadrics $Q_1, Q_2, Q_3$ are selected to be generic such that the eight intersection points are in a generic configuration, the resulting 1/2 Calabi--Yau 3-fold has the Mordell--Weil rank seven.
\par As stated previously, given the equality (\ref{sum 7 in 2.1}) when the 1/2 Calabi--Yau 3-fold develops an $ADE$ singularity the Mordell--Weil rank decreases because the sum of the ranks of the $ADE$ singularity and the Mordell--Weil group always add to 7.  
\par Therefore, when 1/2 Calabi--Yau 3-fold has a rank 7 singularity such as $E_7$ or $A_7$, its Mordell--Weil rank is 0. When 1/2 Calabi--Yau 3-fold has an $ADE$ singularity of rank strictly less than 7, such as $E_6$ and $A_4$, it has a positive Mordell--Weil rank. 

\subsubsection{Construction of Calabi--Yau 3-folds as double covers}
\label{section2.1.2}
\par Further, we turn to the construction of Calabi--Yau 3-folds by taking double covers of 1/2 Calabi--Yau 3-folds on which F-theory is compactified. We consider the double cover of 1/2 Calabi--Yau 3-folds as follows:
\begin{equation}
\label{CY double cover in 2.1}
\tau^2 = F_4(Q_1, Q_2, Q_3).
\end{equation}
In the equation of the double cover (\ref{CY double cover in 2.1}), $F_4$ denotes a homogeneous polynomial of degree 4 in the variables of the quadrics $Q_1, Q_2$, and $Q_3$ \footnote{$F_4$ is a polynomial of degree 4 of the quadrics, and thus it has degree 8 as a polynomial in the coordinates of $\P^3$ as variables. Therefore, the double cover (\ref{CY double cover in 2.1}) is branched over a degree 8 polynomial. The degree is chosen such that the double cover satisfies the Calabi--Yau condition.}. The double cover (\ref{CY double cover in 2.1}) yields an elliptically fibered Calabi--Yau 3-fold. 
\par Since the base surface of 1/2 Calabi--Yau 3-fold is isomorphic to $\P^2$, the equation (\ref{CY double cover in 2.1}) of Calabi--Yau 3-fold implies that the base surface $B$ of Calabi--Yau 3-fold (\ref{CY double cover in 2.1}) is a double cover of $\P^2$ ramified along a quartic curve. As indicated by a result in mathematics, such a surface is known as isomorphic to the del Pezzo surface of degree two \footnote {The blow-up of 7 points on the projective plane $\P^2$ yields this surface. We select a convention to refer to the surface as the degree two del Pezzo surface.}. The middle cohomology group of the second cohomology group of the degree two del Pezzo surface $dP_7$ is known to have eight dimensions, 
\begin{equation}
h^{1,1}(B\cong dP_7) = 8.
\end{equation}
Thus, the number of tensor multiplets arising in F-theory on the elliptically fibered Calabi--Yau 3-fold (\ref{CY double cover in 2.1}) is \cite{MV2}:
\begin{equation}
T=8-1=7.
\end{equation}
\par In contrast to the situation for elliptic K3 surface obtained by gluing a pair of 1/2 K3 surfaces \cite{KRES}, the number of the singular fibers of the resulting Calabi--Yau 3-fold (\ref{CY double cover in 2.1}) is identical to the original 1/2 Calabi--Yau 3-fold. $ADE$ singularity types of the original 1/2 Calabi--Yau 3-fold and the Calabi--Yau 3-fold obtained as double cover are identical. We provide a demonstration that their singularity types are identical in appendix \ref{appendixA}. 

\vspace{5mm}

\par We discuss the Mordell--Weil rank of the resulting Calabi--Yau 3-fold, $Y$, (\ref{CY double cover in 2.1}) as double cover of the original 1/2 Calabi--Yau 3-fold, $X$. The operation of taking a double cover can be regarded as a base change. Therefore, the base change of a solution of the Weierstrass equation yielding a global section of the original 1/2 Calabi--Yau 3-fold $X$ gives a global section of the resulting Calabi--Yau 3-fold $Y$. This implies that the Mordell--Weil group of the original 1/2 Calabi--Yau 3-fold $X$, $MW(X)$, is a subgroup of the Mordell--Weil group of the resulting Calabi--Yau 3-fold $Y$, $MW(Y)$:
\begin{equation}
MW(Y) \supseteq MW(X).
\end{equation}
It follows that the Mordell--Weil rank of the Calabi--Yau 3-fold $Y$ is greater than or equal to the Mordell--Weil rank of the original 1/2 Calabi--Yau 3-fold $X$:
\begin{equation}
\label{inequality of ranks in 2.1}
{\rm rk}\, MW(Y) \ge {\rm rk}\, MW(X).
\end{equation}
\par Further, we expect that the ranks are actually equal for generic values of the parameters of the double cover:
\begin{equation}
\label{MW rank equality in 2.1}
{\rm rk}\, MW(Y) = {\rm rk}\, MW(X),
\end{equation}
The reason for the expectation is explained in appendix \ref{appendixA}.
\par As stated previously, when three quadrics $Q_1, Q_2, Q_3$ are selected as generic, the 1/2 Calabi--Yau 3-fold has Mordell--Weil rank seven. For special choices of the three quadrics, the obtained 1/2 Calabi--Yau 3-folds develop $ADE$ singularity types, and the Mordell--Weil rank decreases. The Mordell--Weil ranks of the 1/2 Calabi--Yau 3-folds range from zero to seven, based on the choices of the three quadrics $Q_1, Q_2, Q_3$. The elliptically fibered Calabi--Yau 3-folds (\ref{CY double cover in 2.1}) obtained as double covers of the 1/2 Calabi--Yau 3-folds have Mordell--Weil ranks greater than or equal to those of the original 1/2 Calabi--Yau 3-folds due to the inequality (\ref{inequality of ranks in 2.1}). $U(1)$ arises in F-theory when the Mordell--Weil rank is strictly positive, and this occurs when the resulting Calabi--Yau 3-fold is obtained as a double cover of the 1/2 Calabi--Yau 3-folds of positive Mordell--Weil ranks. The resulting Calabi--Yau 3-folds (\ref{CY double cover in 2.1}) as double covers of the 1/2 Calabi--Yau 3-folds of Mordell--Weil ranks from one to seven have Mordell--Weil ranks of at least one to seven due to the inequality (\ref{inequality of ranks in 2.1}). From one to seven $U(1)$ gauge group factors form in 6D $N=1$ F-theory on the resulting elliptic Calabi--Yau 3-folds. 

\vspace{5mm}

\par In sections \ref{section2.2}, \ref{section2.3}, \ref{section2.4}, \ref{section2.5}, and \ref{section2.6}, we discuss examples of 6D F-theory models on Calabi--Yau 3-folds as double covers of 1/2 Calabi--Yau 3-folds, with seven, six, five, four, and one $U(1)$ factors as a demonstration of our construction as described above. 

\subsection{Model with U$(1)^7$}
\label{section2.2}
\par When the three quadrics $Q_1, Q_2, Q_3$ are generically chosen, the 1/2 Calabi--Yau 3-fold obtained as a blow-up of $\P^3$ in the eight intersection points of the three quadrics does not have $ADE$ singularity, and it has Mordell--Weil rank seven. Seven sections among the eight sections arising from the blown-up points generate the Mordell--Weil group, as mentioned previously.
\par Given the inequality (\ref{inequality of ranks in 2.1}), the resulting Calabi--Yau 3-fold as double cover of this 1/2 Calabi--Yau 3-fold has the Mordell--Weil rank of at least seven. $U(1)^7$ forms in 6D F-theory on the resulting Calabi--Yau 3-fold. Given that the singularity types of the resulting Calabi--Yau 3-fold and the original 1/2 Calabi--Yau 3-fold are identical as we mentioned in section \ref{section2.1}, F-theory on the resulting Calabi--Yau 3-fold does not have a non-Abelian gauge group factor. 

\subsection{Model with U(1)}
\label{section2.3}
We construct an example of a 6D F-theory model with one $U(1)$ factor by utilizing a 1/2 Calabi--Yau 3-fold with a specific singularity.
\par
\begin{figure}
\begin{center}
\includegraphics[height=10cm, bb=0 0 960 540]{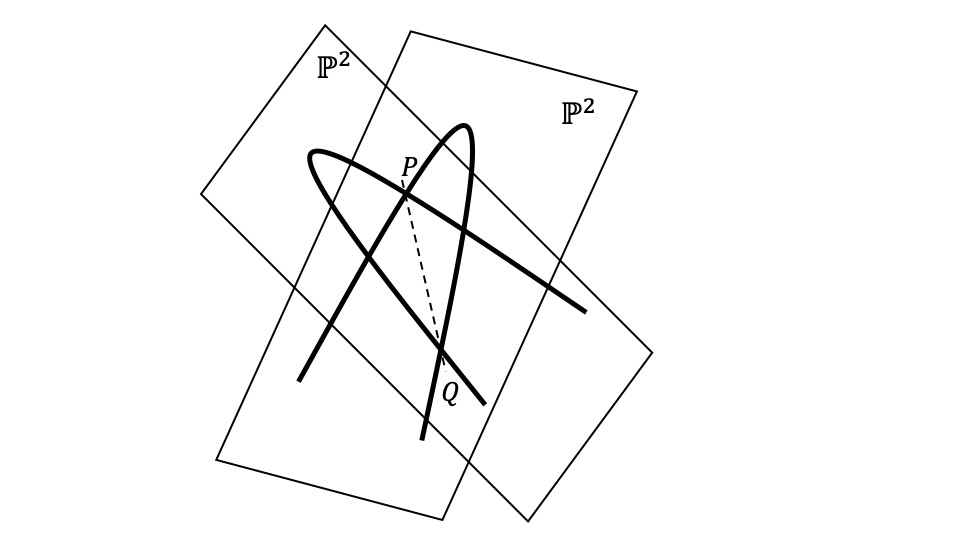}
\caption{\label{figconicsin2.3}Union of two conics intersecting in two points, $P, Q$. Each of the two conics lies inside a hyperplane that is isomorphic to $\P^2$.}
\end{center}
\end{figure}

 We specifically consider the following three quadrics in $\P^3$:
\begin{eqnarray}
\label{three quadrics in 2.3}
Q_1 & = x^2-y^2 \\ \nonumber
Q_2 & = y^2-z^2 \\ \nonumber
Q_3 & = z^2-w^2.
\end{eqnarray}
We used $[x:y:z:w]$ to give coordinates of $\P^3$. The 1/2 Calabi--Yau 3-fold obtained by blowing up the eight intersection points of the three quadrics (\ref{three quadrics in 2.3}) on $\P^3$ actually has the $ADE$ singularity type $A_1^6$. We demonstrate this as follows: we consider the curve $Q_1=0$ in the base surface $\P^2$ under the projection $[Q_1:Q_2:Q_3]$. The condition $Q_1=0$ is as follows:
\begin{equation}
x\pm y=0.
\end{equation}
Therefore, the fiber over a point $[0:a:b]$ of the curve $Q_1=0$ in the base is given by the union of $\{x-y=0, \hspace{1.5mm} b\,Q_2-a\,Q_3=0\}$ and $\{x+y=0, \hspace{1.5mm} b\,Q_2-a\,Q_3=0\}$ in $\P^3$. Using a technique in algebraic geometry, one finds that the union of $\{x-y=0, \hspace{1.5mm} b\,Q_2-a\,Q_3=0\}$ and $\{x+y=0, \hspace{1.5mm} b\,Q_2-a\,Q_3=0\}$ is the union of two conics (embedded inside the intersecting two $\P^2$s) meeting in two points \footnote{This is shown as follows: Given that the hyperplane $x+y=0$ in $\P^3$ yields $\P^2$, $\{x-y=0, \hspace{1.5mm} b\,Q_2-a\,Q_3=0\}$ is isomorphic to a conic in $\P^2$. Similarly, $\{x+y=0, \hspace{1.5mm} b\,Q_2-a\,Q_3=0\}$ is isomorphic to a conic in $\P^2$. The hyperplanes $x+y=0$ and $x-y=0$ yield two distinct $\P^2$s in $\P^3$, and thus the union of $\{x-y=0, \hspace{1.5mm} b\,Q_2-a\,Q_3=0\}$ and $\{x+y=0, \hspace{1.5mm} b\,Q_2-a\,Q_3=0\}$ is isomorphic to the union of two conics embedded inside two meeting $\P^2$s.}. Thus, fiber over the point $[0:a:b]$ in the curve $Q_1=0$ in the base is a type $I_2$ fiber. The image of this is shown in Figure \ref{figconicsin2.3}. Thus, this indicates that the fibers over the curve $Q_1=0$ are type $I_2$ fibers.

\par We also consider three more conics in addition to $Q_1, Q_2, Q_3$, as follows:
\begin{eqnarray}
Q_1+Q_2 & = x^2-z^2 \\ \nonumber
Q_1+Q_2+Q_3 & = x^2-w^2 \\ \nonumber
Q_2+Q_3 & = y^2-w^2.
\end{eqnarray}
The fibers over other five curves, $Q_2=0, \, Q_3=0, \, Q_1+Q_2=0, \, Q_1+Q_2+Q_3=0, \, Q_2+Q_3=0$, are also type $I_2$ based on an argument similar to that we have just used. 
\par We use $l_i$, $i=1, \ldots, 6$, to denote the six curves, $Q_1=0, \, Q_2=0, \, Q_3=0, \, Q_1+Q_2=0, \, Q_1+Q_2+Q_3=0, \, Q_2+Q_3=0$. Subsequently, the arguments that were used above imply that the 1/2 Calabi--Yau 3-fold obtained as the blow-up of $\P^3$ in the base points of the three quadrics (\ref{three quadrics in 2.3}) has singular fibers of type $I_2$ over the six curves $l_i$, $i=1, \ldots, 6$ \footnote{Fibers become singular over the quadric curves in the base surface when the equation of curve consists of two quadratic terms. There are four coordinate variables, namely $x,y,z,w$ and we can select two variables out of the four variables to yield such an equation, e.g. $x^2-y^2$. Therefore, we obtain $\binom{4}{2}=6$ choices, i.e., there are precisely six curves over which fibers become singular in this example. Thus, the six curves, $l_i$, $i=1, \ldots, 6$, yield all curves over which fibers become singular.}, and the discriminant is given as follows:
\begin{equation}
\Delta \sim \prod_{i=1}^6 l_i^2.
\end{equation}
The 1/2 Calabi--Yau 3-fold has the singularity type $A_1^6$. Given the equality (\ref{sum 7 in 2.1}), we deduce that the obtained 1/2 Calabi--Yau 3-fold has Mordell--Weil rank one. 

\vspace{5mm}

\par Double cover (\ref{CY double cover in 2.1}) of the 1/2 Calabi--Yau 3-fold yields an elliptically fibered Calabi--Yau 3-fold, and the $ADE$ singularity types are invariant under this operation as discussed in section \ref{section2.1}. Therefore, the resulting Calabi--Yau 3-fold has the $ADE$ type $A_1^6$. The resulting Calabi--Yau 3-fold has Mordell--Weil rank of at least one due to (\ref{inequality of ranks in 2.1}). F-theory on the Calabi--Yau 3-fold has a gauge group factor as follows:
\begin{equation}
SU(2)^6 \times U(1).
\end{equation}

\subsection{Model with U$(1)^6$}
\label{section2.4}
By applying the result obtained in section \ref{section2.3}, we can construct a 6D F-theory model with six $U(1)$ factors. 
\par We choose the specific quadric $Q_1$:
\begin{equation}
Q_1=x^2-y^2,
\end{equation}
and we choose generic quadrics for $Q_2, Q_3$, to build a 1/2 Calabi--Yau 3-fold. By applying an argument similar to that we have used in section \ref{section2.3}, we learn that the singular fibers over the curve $Q_1=0$ in the base surface $\P^2$ are type $I_2$ fibers. Because two quadrics $Q_2, Q_3$ are generically chosen, fibers become degenerate only over the curve $Q_1=0$. When we use $l$ to denote the curve $Q_1=0$, the discriminant is given as follows:
\begin{equation}
\Delta \sim l^2.
\end{equation}
\par The obtained 1/2 Calabi--Yau 3-fold has the $ADE$ singularity type $A_1$, and it has the Mordell--Weil rank six owing to the equality (\ref{sum 7 in 2.1}). 
\par Because the singularity type of the Calabi--Yau 3-fold obtained by taking double cover (\ref{CY double cover in 2.1}) is identical to the original 1/2 Calabi--Yau 3-fold for generic values of the parameters, the Calabi--Yau 3-fold obtained as double cover (\ref{CY double cover in 2.1}) has the singularity type $A_1$. The resulting Calabi--Yau 3-fold has Mordell--Weil rank at least six. The following gauge group forms in 6D F-theory on the resulting Calabi--Yau 3-fold:
\begin{equation}
SU(2) \times U(1)^6.
\end{equation}

\subsection{Model with U$(1)^5$}
\label{section2.5}
In similar fashions, we can also construct elliptically fibered Calabi--Yau 3-folds with the singularity types $A_1^2$ and $A_1^3$. We discuss the construction of Calabi--Yau 3-folds with the singularity type $A_1^2$ here. The construction of Calabi--Yau 3-folds with the singularity type $A_1^3$ will be discussed in section \ref{section2.6}. 
 \par We consider the quadrics $Q_1, Q_2$ as:
\begin{eqnarray}
Q_1  & = x^2-y^2 \\ \nonumber
Q_2  & = z^2-w^2,
\end{eqnarray}
and we choose the quadric $Q_3$ to be generic. In this situation, we deduce that the 1/2 Calabi--Yau 3-fold obtained as the blow-up of $\P^3$ in the base points of the three quadrics $Q_1, Q_2, Q_3$ has type $I_2$ fibers over the curves $Q_1=0$ and $Q_2=0$ in the base by applying an argument similar to that given in section \ref{section2.3}. When we use $l_1, l_2$ to denote the two curves, the discriminant of the 1/2 Calabi--Yau 3-fold is given as follows:
\begin{equation}
\Delta \sim l_1^2 \cdot l_2^2.
\end{equation}
The resulting 1/2 Calabi--Yau 3-fold has the singularity type $A_1^2$. The 1/2 Calabi--Yau 3-fold has Mordell--Weil rank five owing to the equality (\ref{sum 7 in 2.1}). 
\par The Calabi--Yau 3-fold obtained by taking double cover (\ref{CY double cover in 2.1}) has the singularity type $A_1^2$ for the generic values of the parameters and it has Mordell--Weil rank at least five. The following gauge group forms in 6D F-theory on the resulting Calabi--Yau 3-fold (\ref{CY double cover in 2.1}):
\begin{equation}
SU(2)^2\times U(1)^5.
\end{equation}

\subsection{Model with U$(1)^4$}
\label{section2.6}
We discuss the construction of elliptically fibered Calabi--Yau 3-folds with the singularity type $A_1^3$ here. 
\par We consider the two quadrics $Q_1, Q_2$ given as:
\begin{eqnarray}
Q_1 & = x^2-y^2 \\ \nonumber
Q_2 & = y^2-z^2,
\end{eqnarray}
and we choose the quadric $Q_3$ to be generic. The blow-up of the eight intersection points of the three quadrics $Q_1, Q_2, Q_3$ on $\P^3$ yields a 1/2 Calabi--Yau 3-fold. By using an argument similar to that given in section \ref{section2.3}, we find that the fibers over the curves $Q_1=0$ and $Q_2=0$ in the base surface are type $I_2$ fibers. Furthermore, because $Q_1+Q_2=x^2-z^2$ also splits into linear factors, fibers over the curve $Q_1+Q_2=0$ degenerate to type $I_2$ fibers. 
\par When we denote the three curves, $Q_1=0, Q_2=0, Q_1+Q_2=0$, by $l_1, l_2, l_3$, respectively, then the discriminant of the resulting 1/2 Calabi--Yau 3-fold is given by:
\begin{equation}
\Delta \sim l_1^2\cdot l_2^2 \cdot l_3^2.
\end{equation}
The singularity type of the obtained 1/2 Calabi--Yau 3-fold is $A_1^3$, and we learn that it has the Mordell--Weil rank four owing to the equality (\ref{sum 7 in 2.1}). 
\par The Calabi--Yau 3-fold obtained by taking double cover (\ref{CY double cover in 2.1}) has the singularity type identical to that of the original 1/2 Calabi--Yau 3-fold for the generic values of the parameters. The resulting Calabi--Yau 3-fold (\ref{CY double cover in 2.1}) has the singularity type $A_1^3$. The Calabi--Yau 3-fold has the Mordell--Weil rank at least four. The following gauge group forms in 6D F-theory on the resulting Calabi--Yau 3-fold:
\begin{equation}
SU(2)^3 \times U(1)^4.
\end{equation}

\section{Comparison of possible numbers of exceptional gauge group factors to other 6D F-theory models}
\label{section3}
\par 6D $N=1$ F-theory models in which multiple exceptional gauge group factors arise were constructed in \cite{Kimura1810, Kimura1902}; 6D F-theory models with up to six $E_6$ gauge group factors were constructed in \cite{Kimura1810}, and 6D F-theory models with up to four $E_7$ gauge group factors were constructed in \cite{Kimura1902}. The base surface of Calabi--Yau 3-folds as compactification spaces for these models is isomorphic to $\P^1\times\P^1$, and the number of tensor multiplets that arise in the 6D F-theory models constructed in \cite{Kimura1810, Kimura1902} is one. 
\par Seven tensor multiplets arise in 6D $N=1$ F-theory models on the Calabi--Yau 3-folds that we constructed in this study as discussed in section \ref{section2.1}. When the number of tensor multiplets increases the conditions imposed on the gauge group forming in 6D $N=1$ quantum gravitational theory generally weaken as indicated by the analyses in \cite{KT0910, KMT1008, PT1110} suggest. Based on the tendency, it is natural to expect that there are 6D F-theory models with $T=7$ with the numbers of exceptional $E_6$ and $E_7$ gauge group factors greater than or equal to the numbers of those constructed in \cite{Kimura1810, Kimura1902}. 
\par However, the 1/2 Calabi--Yau 3-folds built in this work can have $E_6$ or $E_7$ singularity type up to one due to the equality (\ref{sum 7 in 2.1}). Given that the singularity type of the elliptic Calabi--Yau 3-fold obtained as double cover (\ref{CY double cover in 2.1}) is identical to that of the original 1/2 Calabi--Yau 3-fold as we will show in Appendix \ref{appendixA}, the construction of elliptically fibered Calabi--Yau 3-folds considered in this study can yield 6D F-theory models with seven tensor fields with $E_6$ or $E_7$ gauge group factor \footnote{6D F-theory models with one $E_7$ factor on Calabi--Yau 3-folds with the base isomorphic to the degree two del Pezzo surface can be constructed by taking double covers (\ref{CY double cover in 2.1}) of the 1/2 Calabi--Yau 3-fold with an $E_7$ singularity.} up to one. Does this mean that the Calabi--Yau 3-folds, on which the expected 6D models with seven tensor fields with the number of $E_6$ and $E_7$ factors equal to or greater than those of the models with one tensor field obtained in \cite{Kimura1810, Kimura1902} are compactified belong to the class that do not allow for splitting into pairs of 1/2 Calabi--Yau 3-folds? A future study can focus on investigating the construction of 6D F-theory models with seven tensor fields with $E_6$ and $E_7$ gauge group factors as many as those with one tensor field constructed in \cite{Kimura1810, Kimura1902}. A physical reasoning suggests the existence of these types of models although the existence is not very clear from our constructions of Calabi--Yau 3-folds in this study. 
\par As demonstrated in \cite{KT0910, KMT1008, PT1110}, the possibilities of the gauge groups that can form in 6D $N=1$ quantum gravitational theories with less than nine tensor fields are finite. With respect to the swampland conditions, it is potentially interesting to estimate the bounds on the number of factors of exceptional gauge groups in six-dimensional $N=1$ quantum gravitational theories with respect to the numbers of tensor multiplets for less than nine tensor multiplets, and this can be explored in a future study.

\section{Concluding remarks and open problems}
\label{section4}
We introduced certain rational elliptic 3-folds that are referred to as 1/2 Calabi--Yau 3-folds, and we constructed elliptically fibered Calabi--Yau 3-folds by taking double covers of the 1/2 Calabi--Yau 3-folds. The construction of elliptically fibered Calabi--Yau 3-folds as discussed in this note provides a systematic method to construct 6D $N=1$ F-theory models with $U(1)$ factors from one to seven \footnote{Some explicit constructions of 6D F-theory models without $U(1)$ gauge group can be found, e.g. in \cite{Kimura1810, Kimura1902}.}. The base surface of the models is isomorphic to the del Pezzo surface of degree two, $dP_7$. Thus, seven tensor multiplets arise in the models. 
\par As discussed in section \ref{section2.1} the sum of the ranks of the Mordell--Weil group and the $ADE$ singularity type of 1/2 Calabi--Yau 3-fold is always seven. Given the property, elliptically fibered Calabi--Yau 3-folds with various Mordell--Weil ranks were obtained by taking double cover of 1/2 Calabi--Yau 3-folds. We consider the double cover of a 1/2 Calabi--Yau 3-fold obtained as blow-up of $\P^3$ in eight points of generic configuration to yield a Calabi--Yau 3-fold of Mordell--Weil rank of at least seven as described in section \ref{section2.2}. $U(1)^7$ arises in 6D $N=1$ F-theory on this Calabi--Yau space. We also presented examples of Calabi--Yau 3-folds with various Mordell--Weil ranks in sections \ref{section2.3}, \ref{section2.4}, \ref{section2.5}, \ref{section2.6}. Six, five, four and one $U(1)$ factors form in 6D F-theory on these Calabi--Yau 3-folds. Appropriately chosen three quadrics yield 1/2 Calabi--Yau 3-folds of the Mordell--Weil ranks two and three in similar fashions. Double covers (\ref{CY double cover in 2.1}) of these yield elliptically fibered Calabi--Yau 3-folds of Mordell--Weil ranks at least two and three. 

\vspace{5mm}

\par Matter fields in 6D $N=1$ F-theory are localized at the intersections of the 7-branes, and the intersections of the 7-branes need to be analyzed to deduce the matter spectra in 6D F-theory. If the Weierstrass equations of the elliptic Calabi--Yau 3-folds constructed in this work are obtained, then the matter spectra and their locations can be deduced from the Weierstrass equations.  
\par The equations of the three quadrics $Q_1, Q_2, Q_3$ uniquely specify the complex structures of the 1/2 Calabi--Yau 3-folds. Thus, the equations of the three quadrics determine the Weierstrass equation of 1/2 Calabi--Yau 3-fold, and the Weierstrass equation of the Calabi--Yau 3-fold as double cover (\ref{CY double cover in 2.1}) can in principle be deduced from the equations of the three quadrics. However, it appears significantly difficult  to determine the Weierstrass equation of the Calabi--Yau 3-fold practically from the equations of the three quadrics $Q_1, Q_2, Q_3$. Exploring the algorithm to explicitly perform the computation can be a direction of future studies. The explicit forms of the rational sections can also be deduced from the Weierstrass equations of the Calabi--Yau 3-folds.
\par We would like to point out that for the examples of Calabi--Yau 3-folds constructed in sections \ref{section2.3}, \ref{section2.4}, \ref{section2.5}, \ref{section2.6}, the locations of the localized matter fields in F-theory can be determined. The locations of the localized matter can be directly deduced from the equations of the three quadrics. We computed the discriminants of the examples of 1/2 Calabi--Yau 3-folds as discussed in sections \ref{section2.3}, \ref{section2.4}, \ref{section2.5}, \ref{section2.6}. The morphism that we described in footnote \ref{footnote13} in Appendix \ref{appendixA} applied to (the curves in) the discriminants of the examples of 1/2 Calabi--Yau 3-folds yields the discriminants (\ref{discriminant CY in appendix}) of the resulting Calabi--Yau 3-folds as double covers (\ref{CY double cover in 2.1}). Via this method, one can explicitly determine the locations of the localized matter fields arising in 6D F-theory on the Calabi--Yau 3-folds constructed in sections \ref{section2.3}, \ref{section2.4}, \ref{section2.5}, \ref{section2.6}. (The Weierstrass equations need to be determined to deduce the matter spectra in F-theory on these Calabi--Yau 3-folds.) 
\par When the equations of the three quadrics, $Q_1, Q_2, Q_3$, are given, the Shioda map \cite{Shioda1989, Shioda, Wazir} on the Calabi--Yau 3-fold as double cover (\ref{CY double cover in 2.1}) can in principle be determined. The explicit forms of the zero-section and the other rational sections, if they are deduced, help to study the detailed structure of the Shioda map.
\par These data can be used to obtain the charges of the hypermultiplets charged under the Abelian gauge group factors.

\vspace{5mm}
   
\par Given that $E_8$ singularity has rank eight, the 1/2 Calabi--Yau 3-folds cannot have $E_8$ singularity due to the equality (\ref{sum 7 in 2.1}). Therefore, the construction of Calabi--Yau 3-folds as double covers of 1/2 Calabi--Yau 3-folds that is introduced in this work does not (at least directly) yield elliptically fibered Calabi--Yau 3-folds with $E_8$ singularity over the base del Pezzo surface of degree two. Determining whether a construction of 6D $N=1$ F-theory models with $E_8$ singularity with seven tensor multiplets is possible is a likely direction of future study.

\section*{Acknowledgments}

We would like to thank Shun'ya Mizoguchi and Shigeru Mukai for discussions. We are also grateful to the referee for improving this manuscript. 

\appendix
\section{Details of the construction}
\label{appendixA}
\par We stated in section \ref{section2.1} that blow-up of $\P^3$ in eight points (that are specified by the intersections of three quadrics) yields 1/2 Calabi--Yau 3-fold, and taking double covers of the 1/2 Calabi--Yau 3-folds yields elliptically fibered Calabi--Yau 3-folds. In this section, we discuss a few details of the structures of the constructions. 

\vspace{5mm}

\par The ranks of the Mordell--Weil group and the $ADE$ singularity type of a 1/2 Calabi--Yau 3-fold $X$ always add to seven as we stated in section \ref{section2.1}. We provide a sketch of a proof of this by using a technique in algebraic geometry. Given that a 1/2 Calabi--Yau 3-fold is the blow-up of eight points on $\P^3$ and this operation increases the Picard number of $\P^3$ by eight \footnote{For complex surfaces and complex manifolds of higher dimensions, the operation of the blow-up in $n$ points increases the Picard number of the manifold by $n$ \cite{Har}.}, the Picard number of 1/2 Calabi--Yau 3-fold is 9 (independent of the complex structure moduli). The Picard number of a 1/2 Calabi--Yau 3-fold does not depend on the configuration of the eight points to be blown up. The fact that the Picard number of any 1/2 Calabi--Yau 3-fold is 9 is used to show the equality (\ref{sum 7 in 2.1}) as follows: the Picard number is the rank of the group called the ``N\'eron--Severi group'', and the ``N\'eron--Severi group'' is the group generated by the divisors (modulo ``algebraic equivalence''). 1/2 Calabi--Yau 3-fold always has the zero-section and the pullback of the line class $\P^1$ in the base surface $\P^2$, independent of the configuration of the blown-up eight points. In addition to the two divisors, 1/2 Calabi--Yau 3-fold generically has seven free sections and for this generic situation, 1/2 Calabi--Yau 3-fold does not have an $ADE$ singularity. In a special situation where 1/2 Calabi--Yau 3-fold has an $ADE$ type singularity, divisors coming from $\P^1$ components of the singular fibers not meeting the zero-section also contribute to the N\'eron--Severi group. The rank of the contribution from the singular fibers is equal to the sum of the numbers of $\P^1$ components (that do not meet the zero-section) in the singular fibers, which is precisely the rank of the $ADE$ singularity. Given that the N\'eron--Severi group is generated by the divisors coming from $\P^1$ components in the singular fibers not meeting the zero-section, the generators of the Mordell--Weil group and the zero-section and the pullback of the line class in the base surface when we add the ranks of the divisors we arrive at the following equality \footnote{Similar formula, known as the Shioda--Tate formula \cite{Shiodamodular, Tate1, Tate2}, holds for elliptic surfaces with a global section.}:
\begin{equation}
\label{equality of ranks in appendix}
\rho(X) = {\rm rk}\, ADE(X) + {\rm rk} MW(X) + 2.
\end{equation}
$\rho(X)$ on the left-hand side denotes the Picard number of 1/2 Calabi--Yau 3-fold $X$ which is nine, and 2 on the right-hand side is the contribution from the zero-section and the pullback of the line class $\P^1$ in the base surface $\P^2$. Thus, plugging the relation $\rho(X)=9$ into the equality (\ref{equality of ranks in appendix}), we obtain the equality (\ref{sum 7 in 2.1}) as follows:
\begin{equation}
7 = {\rm rk}\, ADE(X) + {\rm rk} MW(X).
\end{equation}

\vspace{5mm}

\par The number of the singular fibers of elliptically fibered Calabi--Yau 3-fold $Y$ obtained as a double cover of the original 1/2 Calabi--Yau 3-fold $X$ is identical to that of the original 1/2 Calabi--Yau 3-fold $X$ as stated in section \ref{section2.1}. Therefore, the $ADE$ singularity types of Calabi--Yau 3-fold $Y$ and the original 1/2 Calabi--Yau 3-fold $X$ are identical. We demonstrate the same here. 
\par The structures of singular fibers \footnote{The correspondence between the types of the singular fibers of the elliptic fibration and non-Abelian gauge symmetries formed on the 7-branes in F-theory is discussed in \cite{MV2, BIKMSV}.} of an elliptically fibered Calabi--Yau 3-fold along the codimension one locus in the base surface are essentially identical to those of the elliptic surfaces; Kodaira's classification \cite{Kod1, Kod2} \footnote{Methods to determine the singular fibers of the elliptic surfaces are described in \cite{Ner, Tate}.} applies to singular fibers of elliptic Calabi--Yau 3-folds along the discriminant components. The types of the singular fibers are determined by the vanishing orders of the Weierstrass coefficients and vanishing orders of the discriminant. 
\par We assume that the discriminant of the 1/2 Calabi--Yau 3-fold $X$ is given as follows:
\begin{equation}
\Delta_X \sim \prod_i \, f_i^{n_i}.
\end{equation}
Taking a double cover can be viewed as a base change as mentioned previously, and base change replaces the coordinate variables of the base space. This replaces each of the polynomials $f_i$ with $\til{f}_i$ where $\til{f}_i$ is obtained by plugging cubic polynomials into the variables of the polynomial $f_i$ \footnote{The cubic curves in $\P^2$ have 10 dimensions. Among the same, the cubic curves that pass through the seven points that are blown up to yield the degree two del Pezzo surface have $10-7=3$ dimensions. We denote the basis of these types of cubic curves by $g_1, g_2, g_3$. Then, the base change of $\P^2$ yielding the degree two del Pezzo surface induces morphism from polynomial $f_i$ to polynomial $\til{f}_i$ as follows:
$$
f_i(s,t,u) \rightarrow \til{f}_i(x,y,z)=f_i\, (g_1(x,y,z), g_2(x,y,z), g_3(x,y,z)).
$$
We used $[s:t:u]$ to denote the homogeneous coordinates of $\P^2$, and $[x:y:z]$ to denote the coordinates of degree two del Pezzo surface (with the exception of the blown-up seven points). \label{footnote13}}. Given that the base space is a complex surface and not a complex curve, the resulting polynomial $\til{f}_i$ remains irreducible. This is crucially different from the case of the base change of 1/2 K3 surfaces as discussed in \cite{KRES, Kimura1802, Kimura1903} \footnote{The discriminant of a 1/2 K3 surface is the product of powers of linear factors, and the quadratic base change \cite{SchShio} replaces each linear factor in the discriminant of the 1/2 K3 surface with a quadratic polynomial \cite{KRES}. The base of an elliptic K3 surface is isomorphic to $\P^1$, and thus each quadratic factor in the discriminant of K3 surface that is obtained as the generic quadratic base change of a 1/2 K3 surface splits into two linear factors. Given this property, the number of the singular fibers of the resulting K3 surface is twice that of the original 1/2 K3 surface \cite{KRES}.}. 
\par Thus, the discriminant of the resulting Calabi--Yau 3-fold $Y$ is given as:
\begin{equation}
\label{discriminant CY in appendix}
\Delta_Y \sim \prod_i \, \til{f}_i^{n_i},
\end{equation}
where each $\til{f}_i$ is irreducible. Therefore, the vanishing orders of the discriminant remain invariant after the operation of the base change for generic parameters. Furthermore, an argument similar to that we have just used for the discriminant shows that the vanishing orders of the Weierstrass coefficients also remain unchanged after the base change. We learn from these that the $ADE$ singularity types of Calabi--Yau 3-fold $Y$ and the original 1/2 Calabi--Yau 3-fold $X$ are identical. 

\vspace{5mm}

\par We also stated in section \ref{section2.1.2} that the Mordell--Weil ranks of the original 1/2 Calabi--Yau 3-fold $X$ and the Calabi--Yau 3-fold $Y$ as double cover (\ref{CY double cover in 2.1}) are expected to be identical. We would like to explain this expectation here. 
\par Before we move on to discuss the Mordell--Weil rank of the elliptically fibered Calabi--Yau 3-fold (\ref{CY double cover in 2.1}), we would like to make a remark: we can particularly consider the double cover of the following form to give an elliptically fibered Calabi--Yau 3-fold:
\begin{equation}
\label{deformation CY in 2.1}
\tau^2 = F_4(Q_1, Q_2, Q_3)+\lambda\, G_2(Q_1, Q_2, Q_3)^2.
\end{equation}
$F_4$ denotes a homogeneous polynomial of degree 4 in the quadrics $Q_1, Q_2,$ and $Q_3$ as we stated previously in section \ref{section2.1.2}, and $G_2$ represents a homogeneous polynomial of degree 2 in the quadrics $Q_1, Q_2, Q_3$, given as:
\begin{equation}
G_2=(a\, Q_1+b\, Q_2)(c\, Q_2+d\, Q_3).
\end{equation}
$\lambda$ denotes the parameter. The Calabi--Yau double cover (\ref{deformation CY in 2.1}) can be viewed as a deformation of the Calabi--Yau double cover (\ref{CY double cover in 2.1}), and $\lambda$ parameterizes the deformation. Taking the limit at which $\lambda$ goes to $\infty$, the double cover (\ref{deformation CY in 2.1}) becomes as follows: 
\begin{equation}
\tau^2 = G_2^2.
\end{equation}
This is also equivalent to the following equation:
\begin{eqnarray}
\label{1/2 CY split in 2.1}
\tau = & \pm G_2 \\ \nonumber
       = & \pm (a\, Q_1+b\, Q_2)(c\, Q_2+d\, Q_3).
\end{eqnarray}
The equation (\ref{1/2 CY split in 2.1}) describes a pair of 1/2 Calabi--Yau 3-folds. From this, we find that the limit at which $\lambda$ goes to $\infty$ corresponds to the limit at which elliptically fibered Calabi--Yau 3-fold (\ref{deformation CY in 2.1}) splits into a pair of 1/2 Calabi--Yau 3-folds. Therefore, the elliptically fibered Calabi--Yau 3-fold (\ref{CY double cover in 2.1}) admits a deformation to split into a pair of 1/2 Calabi--Yau 3-folds. This is very analogous to the deformation of elliptic K3 surface splitting into a pair of 1/2 K3 surfaces as described in \cite{KRES}.
\par Now we would like to explain why the Mordell--Weil ranks of the original 1/2 Calabi--Yau 3-fold $X$ and the Calabi--Yau 3-fold $Y$ as double cover (\ref{CY double cover in 2.1}) are expected to be equal using an argument similar to that given in \cite{Kimura1802} as follows: if the Mordell--Weil ranks $MW(Y)$ of the Calabi--Yau 3-folds $Y$ are strictly greater than the Mordell--Weil rank of the original 1/2 Calabi--Yau 3-fold $X$ for generic values of the parameters of taking a double cover, the Calabi--Yau 3-folds $Y$ have a global section -which we call $r$- that does not come from the base change of the Mordell--Weil group $MW(X)$ of the original 1/2 Calabi--Yau 3-fold $X$. When one takes the limit wherein Calabi--Yau 3-fold $Y$ splits into the pair of 1/2 Calabi--Yau 3-folds $X$, global section $r$ splits and this yields global section, which we call $\til{r}$, of each of the pair of 1/2 Calabi--Yau 3-folds $X$. Then, when one considers the base change of the resulting global section $\til{r}$, this yields the global section $r$, which is a contradiction. Thus we expect that the Mordell--Weil ranks of the Calabi--Yau 3-fold $Y$ and the original 1/2 Calabi--Yau 3-fold $X$ are equal for generic values of the parameters of taking a double cover.

\end{document}